%% file: m4_astroph3.tex
\documentclass[structabstract]{aa}  
\usepackage{natbib}
\usepackage{graphicx}
\usepackage{url}

\newcommand{\kms}{$\mathrm {km s}^{-1}$}

\newcommand{\teff}{T$_{\rm eff}$}

\usepackage{txfonts}
\begin{document}
%


   \title{Lithium and sodium in the globular cluster \object{M\,4}}   

   \subtitle{Detection of a Li-rich dwarf star: preservation or
pollution?\thanks{Based on observations taken at ESO VLT Kueyen telescope (Cerro
Paranal, Chile, program: 085.D-0537A).}}

   \author{L. \,Monaco\inst{1}, 
   S. \,Villanova\inst{2}, 
   P. \,Bonifacio\inst{3}, 
   E. \,Caffau\inst{4,3}\thanks{Gliese Fellow}, 
   D. \,Geisler\inst{2},
   G. \,Marconi\inst{1},
   Y. \,Momany\inst{1} 
         \and
  \hbox{H.-G. \,Ludwig\inst{4,3}}
    }

\institute{
European Southern Observatory, Casilla 19001, Santiago, Chile
\and
Universidad de Concepci\'on,
Casilla 160-C, Concepci\'on, Chile
\and
GEPI, Observatoire de Paris, CNRS, Univ. Paris Diderot ; Place Jules
Janssen, 92190 Meudon, France
\and
Zentrum f{\"u}r Astronomie der Universit{\"a}t Heidelberg, Landessternwarte,
K{\"on}igstuhl 12, 69117 Heidelberg, Germany
}

\authorrunning{Monaco et al.}
\mail{lmonaco@eso.org}

\titlerunning{Lithium and sodium in  M\,4}

   \date{Received; accepted}

  \abstract
  %
   {The abundance inhomogeneities of light elements  observed in Globular
Clusters (GCs), and notably the ubiquitous Na-O anti-correlation, are generally
interpreted as evidence that GCs comprise several generations of stars.  There
is an on-going debate as to the nature of the stars which produce the
inhomogeneous elements, and investigating the behavior of several elements is a
way to shed new light on this problem.}
  %
  %
   {We aim at investigating the Li and Na content of the GC M\,4, that is known 
   to have a well defined Na-O anti-correlation.}
  %
  %
   {We obtained moderate resolution (R=17\,000-18\,700) spectra for 91  main
sequence (MS)/sub-giant branch stars of M\,4 with the Giraffe spectrograph  at
the FLAMES/VLT ESO facility. Using model atmospheres analysis we measured
lithium and sodium abundances.}
  %
  %
   {We detect a weak Li-Na anti-correlation among un-evolved MS stars. One star
in the sample, \#\,37934, shows the remarkably high lithium abundance
A(Li)=2.87, compatible with current estimates of the primordial lithium abundance.}
  %
  %
   {The shallow slope found for the Li-Na anti-correlation suggests that lithium
is produced in parallel to sodium. This evidence, coupled with its sodium-rich
nature, suggests that the high lithium abundance of star \#\,37934 may originate
by pollution from a previous generations of stars. The recent detection of a
Li-rich dwarf of pollution origin in the globular cluster NGC\,6397 may also
point in this direction. Still, no clear cut evidence is available against a
possible preservation of the primordial lithium abundance for star  \#\,37934.}


   \maketitle


\section{Introduction}

For many years astronomers have  assumed Galactic Globular Clusters (GCs) as an
example of Single Stellar Population, which constitutes the ideal test-bench for
stellar evolution theories. This view came about mainly by considering the
narrow Red Giant Branches (RGB) which characterize the color-magnitude diagrams
(CMDs) of such systems. A theoretical isochrone of a single metallicity  and
age, typically fits well the observed color-magnitude diagrams. Disturbing
information,   such as the scatter in Na abundances in M3 and M13 noted by 
\citet{Cohen} or the variation in CN or CH strengths
\citep{Norris1981,Norris1983}  or other chemical inhomogeneities which emerged
from spectroscopic analysis,  were typically blamed on mixing processes in the
RGB stars. It was only when it was shown that chemical inhomogeneities  persist
down to Turn-Off stars  \citep[TO,][]{hesser78,cannon98,gratton01}, which in
their centers have temperatures too low to allow the nuclear reactions which may
give rise to the observed anomalies (e.g. the Na enrichment), that it became
necessary to  accept that multiple stellar populations have contributed to the
present day chemical composition of GCs. Different scenarios and possible
polluters from which multiple populations in GCs may originate have been
proposed and, in this respect, gathering information about different chemical
species is crucial to consolidate the observational framework and constrain the
models \citep[see, e.g.,][]{decressin07,ventura10,dercole10,valcarce11}.

Further evidence on the complex nature of GCs has been accumulating in recent
years. Besides the spectacular case of $\omega$~Cen  \citep[see, e.g.,][and
references therein]{villanova07} which is now commonly considered as the remnant
nucleus of an accreted dwarf galaxy, recent high accuracy color-magnitude
diagrams obtained using ACS@HST data, have revealed that several among the most
massive globular clusters (NGC2808, NGC1851, NGC6388, NGC6441) present
photometric evidence for multiple sequences at the main sequence (MS) and/or
sub-giant branch (SGB) level \citep[see][for a review]{P808}. 

Globular Clusters  are among the oldest objects in the universe and, in spite of
the mentioned chemical anomalies, the first generation (FG) of stars in GCs is,
however, easily identified by its low Na content. It is interesting in this
context that all of the GCs studied so far  \citep[$\omega$\,Cen, NGC\,6752,
NGC\,6397, 47\,Tuc, M\,4, see][and references therein]{monaco10,mucciarelli11},
present -- at least in their FG of stars -- a Li content comparable to that of
warm metal-poor halo dwarfs, i.e. the so-called {\it Spite plateau}
\citep[][]{spite82,sbordone10}.  The simplest interpretation of the  {\it Spite
plateau} is that the lithium observed in these old stars is the lithium produced
during the Big Bang \citep{iocco}. If this is the case, however,  there is a
``cosmological lithium problem'', because the level of the  {\it Spite plateau}
is a factor of three to five lower than the primordial lithium predicted by
Standard Big Bang Nucleosynthesis and the baryonic density measured from the
fluctuations of the Cosmic Microwave Background \citep[see, e.g.,][and
references therein]{monaco10,sbordone10}.

In this letter, we report the results of our investigation of the lithium and
sodium abundances in MS/SGB stars in the GC M\,4, based on FLAMES-GIRAFFE/VLT
spectra \citep[][]{pasquini2002}.


\section{Observations and data reduction}

\begin{figure}
  \centering
  \includegraphics[width=9cm]{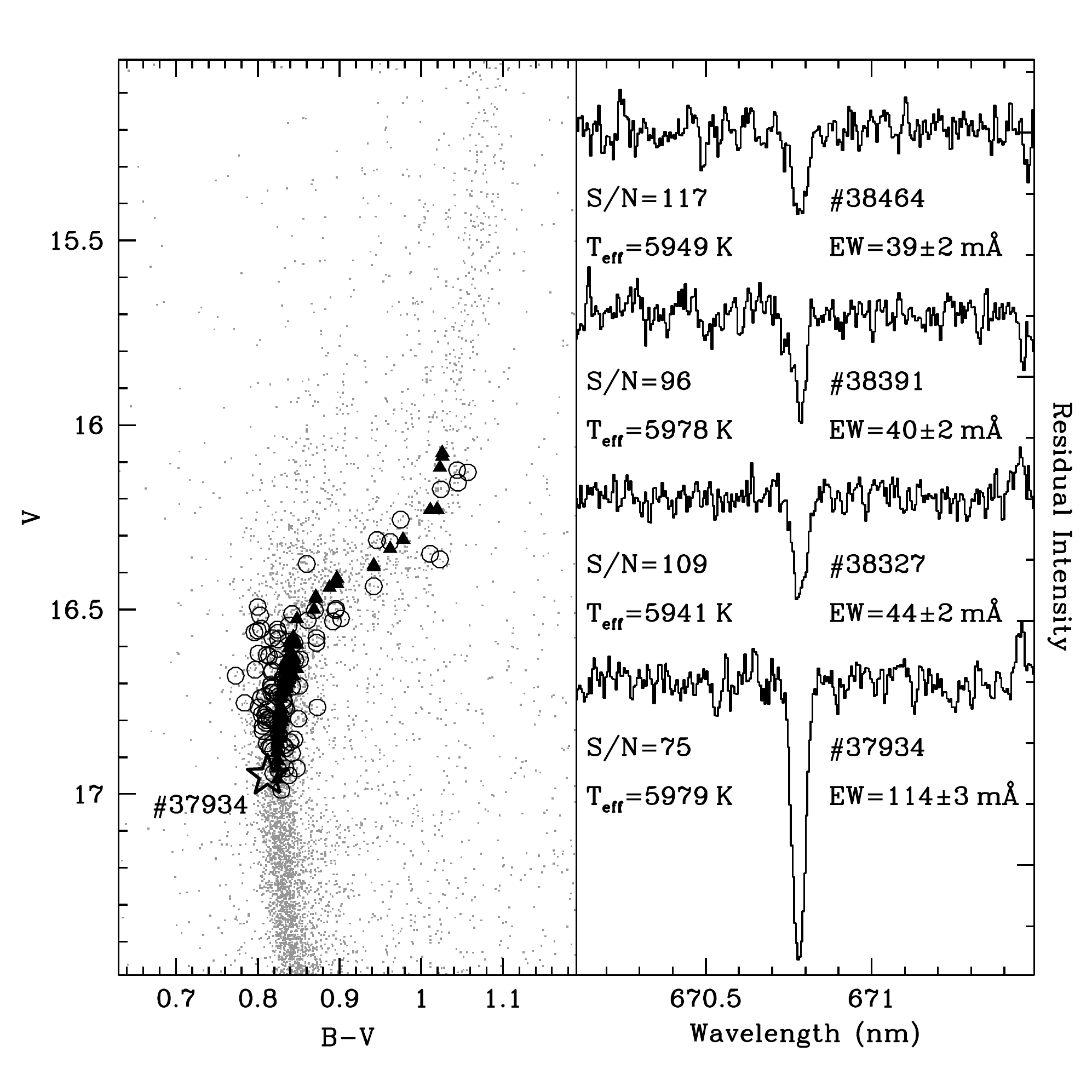} 

     \caption{Left panel: V {\it vs} B-V M\,4 color-magnitude diagram. Target
     stars are marked by open circles. Filled triangles mark the location of the
     target as projected on the cluster mean ridge line. Target \#\,37934 is
     marked by the star symbol. Right panel: summed spectra for a subsample of
     the target stars.} 

	\label{spec}
  \end{figure}

We observed stars along the M\,4 MS and SGB using the FLAMES/GIRAFFE
spectrograph at ESO Paranal (open circles in the left panel of Fig.\ref{spec}).
Observations were conducted in  service mode between April and July 2010 using
the HR12 and HR15N settings. The former covers the Na-D doublet at a resolution
of 18\,700. The HR15N setting  covers the Li\,{\sc i} resonance doublet at
670.8\,nm, as well as  the H$_\alpha$ region at a resolution of 17\,000. The
same plate configuration was used for both settings. Each target was observed
for  $\sim$2.3\,hr and $\sim$10\,hr total integration time in the two settings,
respectively. Frames were processed using version 2.13 of the FLAMES/GIRAFFE
data reduction pipeline\footnote{\url{http://girbldrs.sourceforge.net/}}. 
Ninety-nine medusa fibers were allocated to M\,4 stars, while 15 fibers were
assigned to positions selected for sky subtraction. Sky holes were selected at
similar radial distances from the clusters center as the science targets and the
average of the 15 sky fibers was subtracted from the science spectra.

The standard IRAF\footnote{IRAF is distributed by the National Optical Astronomy
Observatories, which is operated by the association of Universities for Research
in Astronomy, Inc., under contract with the National Science Foundation.} task
{\it fxcor} was employed to measure the stellar radial velocities by
cross-correlating the spectra with synthetic ones of similar atmospheric
parameters. Corrections to the heliocentric system were computed using the IRAF
task {\it rvcorrect} and applied to the observed radial velocities. After being
reduced to rest frame, multiple spectra of the same target were finally
averaged. We obtain spectra with a signal-to-noise ratio (SNR) in the range
72-152 and 19-65 at the Li resonance doublet and the Na-D doublet, respectively.
A sample of the obtained spectra in the Li resonance doublet region is presented
in Fig.\ref{spec} (right panel).

%

We obtain a cluster mean radial velocity and dispersion of
v$_{helio}$=71.4$\pm$0.4\,\kms, $\sigma$=3.7$\pm$0.2\,\kms, after excluding
three stars, deviating more than 3-$\sigma$ from the mean, and five stars
presenting discordant \teff\, as estimated from the  B-V and V-I colors (see
below). These values are in good agreement with the recent measures by
\citet[][]{marino08} and \citet[][]{lovisi10}.


\section{Abundance analysis}

The cluster photometry we adopt in this contribution is based on data collected
with the Wide Field Imager (WFI) mounted at the 2.2m telescope at the La Silla
observatory, which was reduced following \citet[][]{momany03} and was first
corrected for differential reddening following the recipe by
\citet[][]{sarajedini07}. We derived the target stars effective temperatures
(\teff) from the  B-V and V-I colors as projected on the mean cluster ridge line
and using the \citet[][]{alonsodwarf96} calibrations. We adopted for the cluster
a mean reddening of E(B-V)=0.36 \citep[][]{harris96}. The projection we applied
is meant to provide a more robust determination of the targets colors,
particularly given the presence of a significant differential reddening in the
cluster. Indeed, we obtain a good agreement between the \teff\, derived from the
B-V and V-I colors and we eventually adopted their average. Five stars were
excluded from the sample, because of their discrepant positions in the V {\it
vs} B-V and V {\it vs} V-I CMDs, which converted in temperatures derived from
the B-V and V-I colors different by more than 100\,K.

We also estimated the stellar \teff\, by fitting for each star the H$_\alpha$
line with synthetic profiles calculated using the SYNTHE code
\citep[][]{sbordone04,sbordone05,kurucz05}. The two temperature scales are in
excellent agreement, with a mean difference of -6$\pm$5\,K
($\sigma_{\Delta(T_{\rm eff})}$=60\,K). The accuracy of the \teff\,
determination using H$_\alpha$ line fitting is usually, however, of the order of
$\sim$150\,K and is dominated by uncertainties in the continuum normalization
and/or in the correction of the blaze function of the spectrograph
\citep[][]{bonifacio07,sbordone10,monaco10}. In the following we will adopt
150\,K as uncertainty on the \teff\, determination.

Surface gravities were determined with the aid of theoretical isochrones and
vary in the range log\,g=3.48--4.13. In order to determine the stellar
micro-turbulent velocity ($\xi$), stars were first grouped by \teff\, in four
sub-samples. For each sub-sample, a high SNR mean spectrum was generated 
averaging together all the spectra. $\xi$ was then fixed, as usual, by
minimizing the dependence of the derived abundances on the measured equivalent
widths (EWs), for a selected sample of iron lines. The \citet[][]{marino08}
line-list was employed for this purpose and EWs were measured by Gaussian
fitting. The derived trend of $\xi$ as a function of \teff\, was used  to assign
individual $\xi$ values to the target stars. 

We adopted the above atmospheric parameters to calculate proper ATLAS--9 model
atmospheres which we used, along with the MOOG code \citep[][]{sneden73}, to
derive the Fe and Na abundances. For each star, iron abundances were measured
using the two lines at 6995.0\,\AA\, and 6678.0\,\AA. We obtain a mean
[Fe/H]=-1.31$\pm$0.03\,dex (Gaussian dispersion for 71 stars) for MS stars at
\teff$>$5880\,K. This value raises to [Fe/H]=-1.17$\pm$0.03\,dex for the 10
brighter/cooler stars in our sample (\teff$<$5600\,K). This latter value is in
fair agreement (within 0.1\,dex) with previous estimates for RGB/SGB stars
\citep[][]{ivans99,marino08,carretta09,mucciarelli11}. Although compatible
within the errors ($\simeq$0.10-0.15\,dex for both studies), the [Fe/H] we
measure for MS stars is substantially lower than the figure obtained by
\citet[][hereafter M11]{mucciarelli11} for TO stars ([Fe/H]=-1.08). This is
likely due to the different assumptions made about the stellar micro-turbulent
velocities. The possible variation of the iron content with the evolutionary
stage will be analyzed in a forthcoming contribution (Villanova et al., in
preparation).  Na abundances were determined from the EWs of the Na\,D lines at
5889.9\,\AA\, and 5895.9\,\AA\, and applying the corrections tabulated by
\citet[][]{gratton99} for non-LTE (NLTE) effects. We have also asked S.
Andrievsky and S. Korotin to perform some test computations using their sodium
model atom \citep{korM} and version of the multi code \citep[][]{kor1,kor2}, as
done in \citet[][]{andri}. Their computed NLTE corrections are very similar to
those we computed interpolating in the table of \citet[][]{gratton99}.

We derived stellar lithium abundances from the equivalent width (EW) of the 
Li~{\sc i} resonance doublet at 6708~\AA\, using the  \citet[][]{sbordone10}
formula B.1\footnote{IDL routines implementing the fitting formulas are
available at: \url{http://mygepi.obspm.fr/~sbordone/fitting.html}}, which takes
into account 3D (CO5BOLD) and NLTE effects\footnote{We computed ad-hoc 3D-NLTE
line profile for star \#\,37934, whose EW is in the extrapolation regime of the
formula. We found negligible differences (0.003\,dex) with respect to the value
computed with formula B.1.}. We remark that the exact value adopted for the
stellar surface gravity, $\xi$, and iron abundance have minimal impact on
the derived Li abundances, which are, instead dominated by the uncertainty in
the adopted \teff\, and in the measured EWs. We estimated the uncertainty for
the latter according to the Cayrel formula \citep[][]{cayrel88} to be
2-3\,m\AA\,, corresponding to 0.06-0.09\,dex \citep[][]{bonifacio07}. Adding in
quadrature to the error implied by a variation of $\Delta$\teff=$\pm$150\,K,
i.e. an additional 0.09\,dex, we end up with a total uncertainty on the Li
determination of the order of 0.11-0.13\,dex, which we round up to a
conservative 0.15\,dex. We adopt the same error estimate for Fe and Na
abundances as well.

  \begin{figure}
  \centering
  \includegraphics[width=9cm]{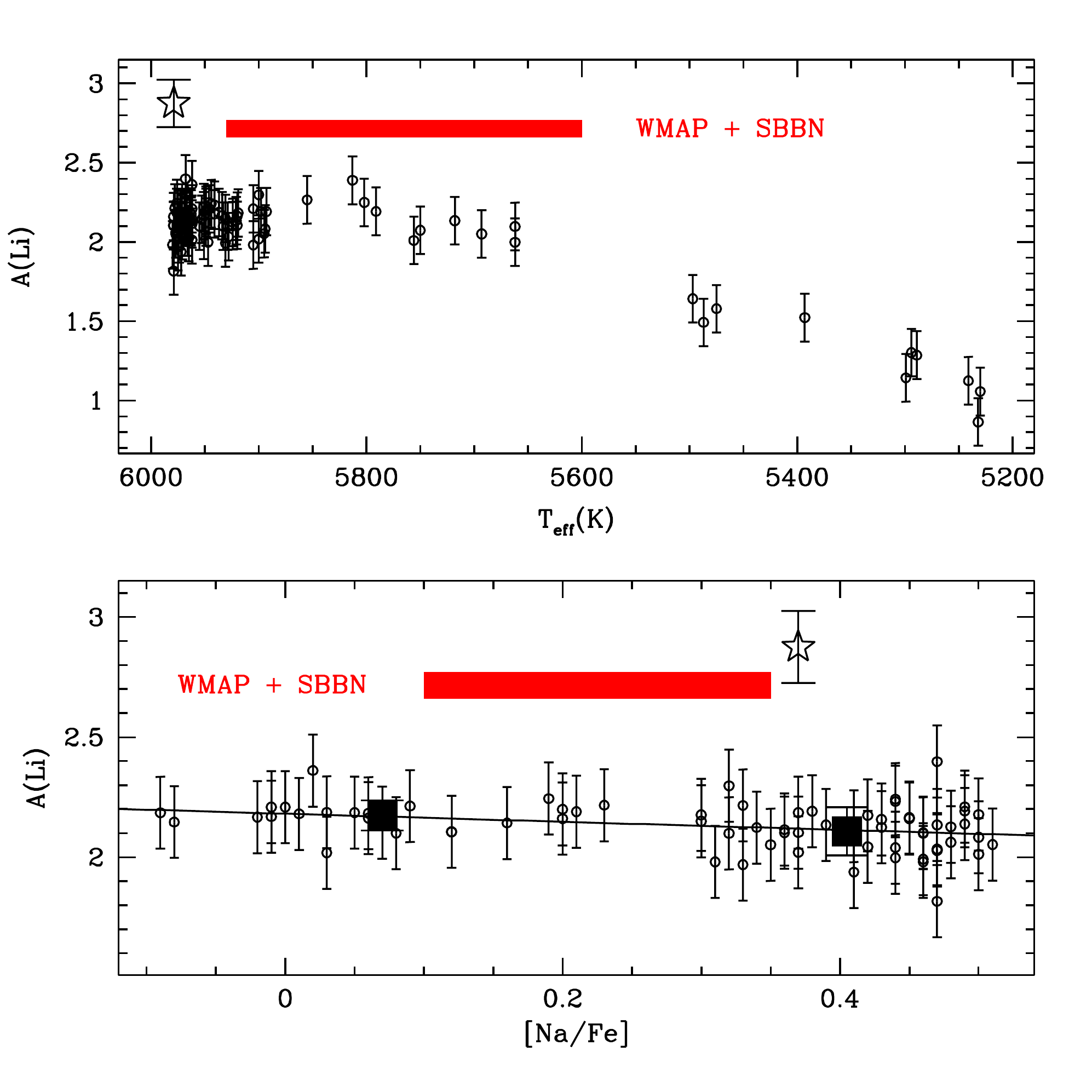}

	\caption{Measured lithium abundances for targets on the main
	sequence/sub-giant branch as a function  of the effective temperature
	(upper panel) and the sodium content (lower panel, stars at
	\teff$>$5880\,K only). The primordial lithium level implied by  WMAP
	measures plus SBBN theory (shaded area) is also marked for reference. In
	the lower panel, filled squares indicate the mean Li content for the
	Na-rich/Na-poor sub-samples. The continuous line is a least square fit
	to the individual data. In both panels, the big open star marks the
	position of star \#\,37934.}

	\label{figteff}

  \end{figure}

%

The upper panel of Fig.\ref{figteff} presents the derived Li abundances as a
function of the stellar \teff. The Li content is relatively constant at the
hottest temperatures and decreases with the temperature for SGB stars. The
observed trend and mean abundances are compatible, within the quoted errors, to
the M11 results. For stars having \teff$>$5880\,K, and excluding \#\,37934 (70
stars), we derive a mean lithium abundance of $<$A(Li)$>$=2.13$\pm$0.09\,dex
(Gaussian dispersion). No correlation of the lithium abundance with temperature
is present among the stars of this sub-sample. M11 derived
$<$A(Li)$>$=2.30$\pm$0.10\,dex for the 35 TO stars in their sample having
\teff$>$5900\,K. The two values are compatible with each other, particularly
considering the different \teff\, scales adopted, the spectral SNR, and the
different CMD regions sampled. The iron content has an impact on the Li
abundances derived through formula B.1. However, an increase of $\Delta$
[Fe/H]=+0.25\,dex would correspond to a negligible decrease of the lithium
abundance of $\Delta$ A(Li) = -0.003\,dex.


\section{Discussion}

Our main results are summarized in the lower panel of Fig.\ref{figteff}, where
we restrict our analysis to MS stars only (\teff$>$5880\,K, 71 stars). The
cluster stars display a very mild Li-Na anti-correlation. Compared to NGC\,6752,
in which a spread of 0.6\,dex in [Na/Fe] corresponds to a spread of 0.6\,dex in
A(Li) \citep{pasquini05}, the anti-correlation in M\,4  is hardly detectable,
with a spread of 0.1\,dex in A(Li) corresponding to a spread of 0.6\,dex in
A(Na).  The sample has, however, a Spearman rank correlation coefficient of
C$_S$=-0.34, corresponding to a probability of 0.002 that the observed
correlation is, in fact, spurious. This conclusion is also supported by an
additional non-parametric test (Kendall's tau), as well as by parametric ones. 
We consider the results of the non-parametric tests as more robust since
parametric tests hinge on the correctness of the underlying model -- the assumed
functional relationship between the variables.  On request of the referee, we
nevertheless also provide the outcome of the parametric tests (see appendix). 
Notice, however, that we detected a very shallow slope and the conclusion we
draw below are thus equally applicable to the case of a non detection of a Na-Li
anti-correlation.

It would be tempting to interpret the detected anti-correlation by a simple
pollution scenario, by which varying quantities of Na-rich/Li-poor material are
mixed with ``pristine'' material. In the case of NGC\,6752, it was already
pointed out by \citet[][]{shen10}, however, that such a scenario would imply a
slope of one in the A(Li)-A(O) correlation. The fact that the observed slope is
significantly different from unity implies that the material is not totally
depleted in Li, requiring a production of some Li along with Na. In the same
vein, we may argue that the fact that clusters like NGC\,6752 and M\,4 show
similar Na-O anti-correlations \citep[][]{marino08} but different Li-Na
anti-correlations, suggests that these anti-correlations arise from the
operation of different nuclear reactions which take place at different places,
and possibly at different times. If Li were simply destroyed along with Na
production, we should always find the same slope of the Li-Na anti-correlation.
On the contrary, we need to postulate that Li is produced with different
efficiency in different clusters, to explain the absence of a slope in NGC\,6397
\citep[][]{lind09}, the mild slope in M\,4 and the marked slope in NGC\,6752. 
Notice also that the observed slopes do not change in direct correlation with
the cluster metallicities. Indeed, \citet[][hereafter D10]{dorazi10a} and M11
reached similar conclusions based on the non-detection of a Li-Na/Li-O
anti-correlation/correlation among a sample of RGB/TO stars. The two groups of
Na-poor/Na-rich stars (arbitrarily separated at [Na/Fe]=0.25) have different
dispersion around the mean of 0.06 and 0.10, respectively, to be compared with
the 0.06 and 0.14 reported by D10.

\citet{dercole10} presented a class of models capable of reproducing the Na-O
and Mg-Al anti-correlations, using  AGB stars of the first generation as
``polluters''. In their model for M\,4, lithium is produced in parallel to the
sodium production. These models are quite successful, but depend on several
parameters and thus have the possibility of producing the different correlations
found in the different clusters. A necessary and key ingredient is the dilution
with pristine gas \citep{dercole11}. A different class of models, in which the
``polluters'' are fast rotating massive stars (FRMS) has been presented by
\citet{decressin07,decressin10}. In this case there is no lithium production and
lithium is purely destroyed in the massive stars. The observed lithium is the 
result of dilution of Li-free matter with pristine material. In a case like that
of M\,4 the material must clearly be almost pristine, otherwise the variation in
Li would be much larger. The observations of the simultaneous, significant
variation in the Na content might pose a problem to the viability of such models
for M\,4. Notice, however, that the recent results by \citet[][]{villanova11}
support FRMS as likely polluters from which the second generation of stars
formed in M\,4 \citep[see also][]{yong08,lind11}.

Figure\,\ref{figteff} shows another remarkable fact: there is one star,
\#\,37934, that displays a lithium abundance which is significantly higher than
that of the others. This star has in fact a lithium content which is compatible
with the primordial lithium abundance, as derived from the baryonic density
deduced from the fluctuations of the Cosmic Microwave Background and Standard
Big Bang Nucleosynthesis \citep[][shaded area in Fig.\ref{figteff}]{cyburt}. The
strong Li line of this star can also be appreciated in Figure\,\ref{spec}. We
have checked individual frames and the strong Li line is detected in all of
them. Observations were conducted over a long period, which implies observations
with different plates and at different geocentric velocities. Hence, different
fibers were used and the star spectrum moved on the CCD. This guarantees that
the detection is not spurious. It is natural to pose the question whether it is
a case of preservation or production.  It is currently accepted that the
constant Li abundance displayed by warm metal-poor stars, the so-called {\em
Spite plateau} \citep{spite82}, falls short by a factor of three to five of the
primordial lithium abundance  \citep[see][and references therein]{sbordone10}.
With the exception of star \#\,37934, the picture we are facing in M\,4 is in
fact similar to  what we see among field stars \citep[the mild Li-Na
anti-correlation is hardly significant in this context, see also M11
and][]{monaco10}. Notice that the \citet[][]{robin03} model predicts no Galactic
stars over a one square degree area in the M\,4 line of sight with color,
magnitude, radial velocity and metallicity similar to \#\,37934, within
$\pm$0.1\,mag, $\pm$0.5\,mag, $\pm$3\,$\sigma$ and $\pm$0.5\,dex, in each
variable, respectively.

Two metal-poor dwarf stars that lie significantly above the Spite plateau are
known, BD +23$^\circ$ 3912 \citep{BM97}  and HD\,106038 \citep{A06}. These stars
have lithium abundance differences with respect to the cosmological value
similar as \#\,37934, but at abundances lower than the cosmological value
(A(Li)=2.72). The 1D-NLTE lithium abundances of the three stars are A(Li)=2.83,
2.59 and 2.48  for \#\,37934, BD +23$^\circ$ 3912 and HD\,106038, respectively.
Notice that adopting the recent temperature calibrations by \citet{gh09}, we
obtain a temperature 116\,K hotter for star \#37934 and a corresponding 1D-NLTE
abundance of A(Li)=2.92. Stars in this metallicity range could already start to
feel the effect of the Galactic enrichment in lithium \citep{romano99,romano03}.
The remarkable difference between star \#\,37934 and BD\,+23\,3912 or HD 106038,
is the gap  in Li abundance with respect to the reference population: it is
$\sim 0.7$\,dex for star \#\,37934 but only $\sim 0.3$\, dex for the other two
stars. HD 106038 displays other chemical peculiarities,  such as overabundance
of Be \citep{rodolfo}, Si, Ni, Y and Ba. \citet{rodolfo} invoke a hypernova to
explain such peculiarities,  without however, explaining how a star may form
from the undiluted hypernova ejecta. It is clear that such an exotic explanation
is not applicable to the case of Globular Cluster stars, since the hypernova
explosion would expel all gas from the Cluster, thus stopping star-formation.

Also in NGC\,6752 and 47\,Tuc \citep[][]{shen10,dorazi10b} there is one star,
which, within errors, has a Li abundance compatible with the primordial
abundance. In those cases, however,  the star lies along a well defined Li-O
correlation so that there is a continuity of Li abundances from the highest to
the lowest. Here the situation is totally different, as there is no star with a
Li content within a factor of 3 of that of \#\,37934. 

The nature of this star can be interpreted in different ways. It is currently a
matter of debate whether the Spite Plateau originate from a depletion mechanism
which is experienced by the stars in a uniform way. Star \#\,37934 might have
escaped this depletion which, instead, experienced all the other stars we
sampled. In this case the high Li abundance of \#\,37934 would represent a case
of preservation of the primordial Li abundance. On the other hand, this star
might have followed the same fate of the other stars but might have been created
with a higher Li abundance. Its high lithium content would then constitute a
case of pollution.

We believe it is hard to rule out any of the two interpretations. In the first
case we should identify a mechanism which uniformly depletes lithium in all the
cluster stars and  another mechanism which suppresses this depletion for star 
\#\,37934.  In the second case, although it is relatively easy to envisage Li
production mechanisms, it is not so easy to imagine mechanisms which would allow
this single star to be so much more lithium-rich than  the others. The small
slope of the Li-Na anti-correlation supports Li production taking place in
parallel to Na production. Together with the fact that star \#\,37934 is Na rich
(and thus a ``second generation star''), this provides circumstantial evidence
that favors the second scenario. Additionally, the measures on the single
exposures suggest a possible small radial velocity variation at the level of
$\sim$1.5\,\kms over a period of $\sim$75 days, which, in turn, may support the
pollution scenario from a (now evolved) companion AGB star. An extremely Li-rich giant star
has been detected in the Globular Cluster M\,3 \citep{Kraft}. Such Li-rich
phases are of short duration, however if the star has an unevolved companion,
and transfers Li-rich material onto its atmosphere during this phase, the
companion may preserve this Li-rich material. That a local production of Li
should result in a lithium abundance matching the primordial  lithium abundance
would, however, be a remarkable coincidence. Recently, \citet[][]{koch11} have
reported the detection of a Li-rich dwarf (A(Li)=4.03) in the globular cluster
NGC\,6397. This certainly constitute a case of pollution and reinforces the
pollution scenario also for star \#\,37934.

Whichever its origin, star \#\,37934 shows that old, metal-poor, un-evolved
stars with the cosmic lithium abundance do exist. We may expect to find other
examples of such Li-rich stars in M\,4 and other clusters. To the extent that a
part of the Halo field population may have been formed from stars lost by GCs,
we should be able to find similar stars also in the field population. 

\begin{acknowledgements}

We are grateful to S. Andrievsky and S. Korotin for performing some test NLTE
computations for Na. PB acknowledges support from CNRS INSU through PNPS and
PNCG grants.

\end{acknowledgements}

\appendix

\section{The Na-Li anticorrelation}

On request of the referee we provide the results of our parametric tests in
this section. The familiar methodology underlying linear regression makes it
appealing, however, the reader should be aware that we base our conclusions
about the reality of the anti-correlation on the non-parametric tests.

We select  only stars with \teff $> 5880$ and A(Li)$< 2.6$ thus excluding star
\# 37934 (inclusion of the star does not change the results in any significative
way). We perform a straight line fit taking into account errors in both $x$ and
$y$ using routine {\tt fitexy} \citep{numrec}.

We assumed a conservative error of 0.15 dex for the abundances measured in this
contribution. 0.15\, dex is an appropriate ``external'' error, however in the
context of the Na-Li slope what is relevant is the ``internal'' error, that is
certainly smaller. We assumed an error of 0.09\,dex, guided by the R.M.S of the
linear fits we performed.  Incidentally, this is consistent with the estimated
Li abundance variation due to uncertainty in the EW measures. 

If we use the LTE Na abundance the slope is $-0.30\pm 0.09$ corresponding to a
3.3$\sigma$ detection, the $\chi^2 = 647$ corresponding to a probability of the
fit of 0.59. The root mean square deviation around the fitted line is 0.09\,dex.
Using NLTE Na the slope is $-0.22\pm 0.07$ corresponding to a 3.1$\sigma$
detection. The $\chi^2 = 661$ corresponding to  a probability of 0.54 and the 
root mean square deviation around the fitted line is still 0.09\, dex.

In summary, the parametric fits fully support the conclusion of the
non-parametric tests, suggesting that an anti-correlation between Na and Li 
exists. Statistical considerations of the goodness-of-fit suggest that the
``internal'' errors on the abundances are rather of the order of 0.09\,dex,
while 0.15\, dex is a good estimate of the ``external'' errors.

\Online

\input{onlinecatalog1.tex}
\input{onlinecatalog2.tex}

\end{document}

%% file: onlinecatalog1.tex
\begin{table*}
\caption{Basic data for MS/SGB stars studied in this paper.}
\begin{center}								     
\begin{tabular}{r|cc|ccc|cc|ccc}   					     
\hline									     
ID & RA  & Dec & \teff (K) & log g & $\xi$ (\kms) & A(Fe) & A(Na) & A(Li) & EW (m\AA) & S/N  \\ 	
\hline								             
   506 &16:24:13.10 &-26:23:34.70 &5933 &4.00  &  1.72 &6.16 &5.30 &2.10 & 34 & 87\\
  1024 &16:24:12.60 &-26:22:28.50 &5924 &3.99  &  1.71 &6.19 &5.37 &2.10 & 34 & 72\\
  7746 &16:23:29.61 &-26:24:09.10 &5905 &3.96  &  1.71 &6.19 &5.50 &2.21 & 44 &125\\
  8029 &16:23:48.66 &-26:23:35.10 &5921 &3.98  &  1.71 &6.18 &5.39 &2.13 & 37 &121\\
  8332 &16:23:47.95 &-26:22:50.00 &5928 &3.99  &  1.72 &6.17 &5.46 &2.03 & 30 &107\\
  8405 &16:23:39.36 &-26:22:40.20 &5900 &3.95  &  1.71 &6.17 &5.31 &2.30 & 52 &121\\
  8784 &16:23:36.47 &-26:21:40.20 &5951 &4.02  &  1.72 &6.18 &4.98 &2.17 & 38 &104\\
  9139 &16:23:49.05 &-26:20:40.90 &5905 &3.95  &  1.71 &6.13 &5.41 &1.98 & 28 &110\\
  9254 &16:23:41.92 &-26:20:18.90 &5475 &3.69  &  1.06 &6.28 &5.44 &1.58 & 25 &139\\
  9280 &16:23:44.05 &-26:20:13.10 &5976 &4.10  &  1.73 &6.21 &4.94 &2.18 & 38 & 95\\
  9505 &16:23:48.90 &-26:19:26.80 &5920 &3.97  &  1.71 &6.21 &5.15 &2.11 & 35 & 96\\
 13512 &16:23:05.69 &-26:22:49.70 &5967 &4.06  &  1.72 &6.17 &5.46 &2.13 & 35 & 94\\
 13810 &16:23:03.38 &-26:21:58.30 &5947 &4.01  &  1.72 &6.17 &5.43 &2.00 & 27 & 88\\
 28521 &16:23:13.71 &-26:38:57.30 &5963 &4.05  &  1.72 &6.20 &5.32 &2.18 & 38 & 89\\
 28814 &16:23:11.56 &-26:38:13.70 &5898 &3.95  &  1.71 &6.19 &5.50 &2.19 & 43 &113\\
 30253 &16:23:13.10 &-26:35:02.60 &5230 &3.48  &  1.04 &6.36 &5.51 &1.06 & 13 &131\\
 30699 &16:23:15.92 &-26:34:06.30 &5791 &3.86  &  1.69 &6.19 &5.06 &2.19 & 50 &152\\
 30887 &16:23:15.82 &-26:33:42.40 &5931 &3.99  &  1.72 &6.23 &5.35 &2.15 & 38 &127\\
 31190 &16:23:14.43 &-26:33:06.90 &5971 &4.07  &  1.72 &6.14 &5.40 &2.04 & 28 & 73\\
 31634 &16:23:12.07 &-26:32:15.20 &5241 &3.50  &  1.04 &6.34 &5.28 &1.12 & 15 &120\\
 31931 &16:23:12.77 &-26:31:44.60 &5299 &3.57  &  1.04 &6.34 &5.50 &1.14 & 14 &148\\
 32111 &16:23:13.32 &-26:31:25.10 &5974 &4.08  &  1.72 &6.18 &5.37 &2.10 & 32 &126\\
 32349 &16:23:11.46 &-26:31:00.60 &5294 &3.57  &  1.04 &6.31 &5.32 &1.30 & 19 &144\\
 32890 &16:23:15.02 &-26:30:01.90 &5693 &3.79  &  1.55 &6.23 &5.16 &2.05 & 44 &138\\
 32992 &16:23:13.75 &-26:29:50.10 &5975 &4.08  &  1.72 &6.19 &5.43 &2.17 & 37 & 89\\
 33132 &16:23:10.36 &-26:29:30.70 &5979 &4.13  &  1.73 &6.17 &5.19 &2.16 & 36 & 99\\
 33215 &16:23:15.08 &-26:29:20.00 &5979 &4.10  &  1.73 &6.14 &5.42 &2.10 & 32 &101\\
 33303 &16:23:08.57 &-26:29:08.30 &5750 &3.83  &  1.68 &6.19 &5.45 &2.07 & 43 &128\\
 33349 &16:22:50.71 &-26:29:02.80 &5975 &4.09  &  1.72 &6.21 &5.36 &1.97 & 24 & 83\\
 33462 &16:23:16.48 &-26:28:49.20 &5662 &3.77  &  1.48 &6.22 &5.10 &2.10 & 51 &102\\
 33518 &16:23:13.79 &-26:28:42.30 &5497 &3.70  &  1.11 &6.29 &5.37 &1.64 & 27 &151\\
 33548 &16:23:07.86 &-26:28:38.00 &5289 &3.57  &  1.04 &6.29 &5.12 &1.29 & 19 &123\\
 33744 &16:23:13.91 &-26:28:13.30 &5894 &3.95  &  1.71 &6.16 &5.48 &2.08 & 35 &129\\
 33789 &16:23:15.40 &-26:28:08.00 &5813 &3.87  &  1.69 &6.17 &5.11 &2.39 & 69 &140\\
 33879 &16:23:10.13 &-26:27:55.90 &5977 &4.09  &  1.73 &6.20 &5.37 &2.05 & 29 &100\\
 34383 &16:22:49.09 &-26:26:46.00 &5962 &4.05  &  1.72 &6.21 &5.05 &2.36 & 54 &106\\
 35614 &16:23:28.79 &-26:40:27.10 &5953 &4.03  &  1.72 &6.22 &5.11 &2.14 & 36 &110\\
 35854 &16:23:40.92 &-26:40:12.10 &5969 &4.06  &  1.72 &6.18 &5.03 &2.19 & 38 &105\\
 36554 &16:23:40.47 &-26:39:19.70 &5919 &3.97  &  1.71 &6.18 &5.06 &2.18 & 41 &119\\
 36853 &16:23:37.06 &-26:38:58.60 &5964 &4.05  &  1.72 &6.20 &5.23 &2.19 & 39 & 81\\
 36867 &16:23:19.42 &-26:38:57.90 &5920 &3.98  &  1.71 &6.26 &5.14 &2.16 & 39 &103\\
 37028 &16:23:25.09 &-26:38:45.80 &5893 &3.94  &  1.71 &6.21 &5.41 &2.19 & 43 &119\\
 37156 &16:23:27.21 &-26:38:38.40 &5931 &3.99  &  1.72 &6.24 &5.52 &1.99 & 27 &114\\
 37166 &16:23:35.88 &-26:38:37.90 &5393 &3.65  &  1.03 &6.35 &5.36 &1.52 & 26 &141\\
 37222 &16:23:20.13 &-26:38:34.60 &5895 &3.95  &  1.71 &6.16 &5.49 &2.05 & 33 &111\\
 37934 &16:23:23.71 &-26:37:42.30 &5979 &4.11  &  1.73 &6.16 &5.35 &2.87 &114 & 75\\
 38327 &16:23:30.82 &-26:37:15.90 &5941 &4.01  &  1.72 &6.18 &5.44 &2.23 & 44 &109\\
 38391 &16:23:45.59 &-26:37:11.20 &5978 &4.10  &  1.73 &6.25 &5.16 &2.21 & 40 & 96\\
 38464 &16:23:36.34 &-26:37:06.20 &5949 &4.02  &  1.72 &6.13 &5.32 &2.19 & 39 &117\\
 38485 &16:23:46.46 &-26:37:04.30 &5951 &4.02  &  1.72 &6.13 &5.37 &2.04 & 29 &101\\
 38547 &16:23:28.74 &-26:37:00.10 &5924 &3.98  &  1.71 &6.23 &5.39 &2.12 & 36 &112\\
 38548 &16:23:27.21 &-26:37:00.10 &5487 &3.69  &  1.09 &6.34 &5.47 &1.49 & 20 &117\\
 38573 &16:23:22.03 &-26:36:57.80 &5968 &4.06  &  1.72 &6.16 &5.45 &2.40 & 57 & 98\\
 38597 &16:23:40.83 &-26:36:56.10 &5934 &4.00  &  1.72 &6.17 &5.44 &2.17 & 39 &109\\
 39344 &16:23:24.33 &-26:36:15.30 &5962 &4.04  &  1.72 &6.19 &5.51 &2.01 & 27 & 95\\
 39639 &16:23:23.37 &-26:36:02.20 &5955 &4.03  &  1.72 &6.12 &5.40 &2.10 & 33 &109\\
 39862 &16:23:21.73 &-26:35:52.80 &5662 &3.77  &  1.48 &6.24 &5.43 &2.00 & 42 &138\\
 43701 &16:23:46.92 &-26:33:20.40 &5969 &4.06  &  1.72 &6.16 &5.43 &2.16 & 36 &126\\
 56218 &16:23:25.02 &-26:27:19.70 &5718 &3.81  &  1.61 &6.20 &5.29 &2.13 & 50 &126\\
 57062 &16:23:46.57 &-26:26:54.80 &5962 &4.04  &  1.72 &6.16 &5.14 &2.14 & 35 &103\\
 57261 &16:23:23.78 &-26:26:48.90 &5232 &3.49  &  1.04 &6.35 &5.51 &0.86 &  8 &115\\
 57631 &16:23:24.06 &-26:26:37.30 &5756 &3.83  &  1.68 &6.25 &5.52 &2.01 & 37 &122\\
 57794 &16:23:27.07 &-26:26:32.20 &5979 &4.11  &  1.73 &6.17 &5.11 &2.11 & 32 &104\\
 58082 &16:23:22.16 &-26:26:22.40 &5949 &4.02  &  1.72 &6.21 &5.23 &2.20 & 40 &114\\
 58089 &16:23:20.59 &-26:26:22.20 &5972 &4.07  &  1.72 &6.24 &4.98 &2.15 & 35 &104\\
\hline		
\end{tabular}	
\end{center}	
\end{table*}	

%% file: onlinecatalog2.tex
\addtocounter{table}{-1}
\begin{table*}
\caption{continued.}
\begin{center}								     
\begin{tabular}{r|cc|ccc|cc|ccc}   					     
\hline									     
ID & RA  & Dec & \teff (K) & log g & $\xi$ (\kms) & A(Fe) & A(Na) & A(Li) & EW (m\AA) & S/N  \\ 	
\hline								             
 58440 &16:23:43.05 &-26:26:10.50 &5979 &4.11  &  1.73 &6.18 &5.47 &1.82 & 18 &107\\
 58482 &16:23:35.34 &-26:26:09.10 &5968 &4.06  &  1.72 &6.23 &5.48 &2.12 & 34 &125\\
 58580 &16:23:47.24 &-26:26:05.80 &5965 &4.06  &  1.72 &6.19 &5.48 &2.03 & 28 &105\\
 58671 &16:23:30.37 &-26:26:03.00 &5980 &4.11  &  1.73 &6.22 &5.35 &1.98 & 25 &120\\
 58759 &16:23:42.47 &-26:25:59.70 &5855 &3.91  &  1.70 &6.21 &5.34 &2.27 & 52 &142\\
 59664 &16:23:49.67 &-26:25:26.60 &5967 &4.07  &  1.72 &6.20 &5.01 &2.17 & 37 &112\\
 60462 &16:23:23.79 &-26:24:54.80 &5973 &4.07  &  1.72 &6.18 &5.01 &2.18 & 38 & 95\\
 60508 &16:23:21.25 &-26:24:53.00 &5946 &4.01  &  1.72 &6.25 &5.06 &2.21 & 41 &119\\
 60599 &16:23:22.44 &-26:24:49.40 &5976 &4.09  &  1.72 &6.26 &5.27 &2.24 & 43 &119\\
 63647 &16:24:07.51 &-26:35:36.50 &5964 &4.05  &  1.72 &6.17 &5.48 &2.14 & 35 & 91\\
 63900 &16:23:59.44 &-26:35:06.00 &5966 &4.06  &  1.72 &6.19 &5.49 &2.06 & 30 & 89\\
 64263 &16:24:05.88 &-26:34:24.20 &5951 &4.02  &  1.72 &6.17 &5.22 &2.22 & 42 & 96\\
 65049 &16:24:09.05 &-26:32:56.80 &5972 &4.08  &  1.72 &6.16 &5.41 &2.16 & 36 & 89\\
 65318 &16:24:01.82 &-26:32:24.60 &5937 &4.00  &  1.72 &6.16 &5.03 &2.19 & 40 &107\\
 65824 &16:24:11.26 &-26:31:28.80 &5802 &3.87  &  1.69 &6.17 &5.29 &2.25 & 55 &113\\
 65935 &16:24:03.79 &-26:31:17.50 &5963 &4.05  &  1.72 &6.17 &5.40 &2.13 & 34 &135\\
 65939 &16:24:21.11 &-26:31:17.30 &5900 &3.95  &  1.71 &6.14 &4.99 &2.02 & 30 &103\\
 66076 &16:24:08.33 &-26:31:02.10 &5941 &4.01  &  1.72 &6.14 &5.46 &2.18 & 39 &115\\
 66136 &16:24:09.44 &-26:30:55.80 &5944 &4.01  &  1.72 &6.19 &5.45 &2.24 & 44 &118\\
 66304 &16:24:20.05 &-26:30:36.50 &5974 &4.08  &  1.72 &6.17 &5.36 &2.02 & 27 & 91\\
 66558 &16:24:09.93 &-26:30:08.90 &5962 &4.05  &  1.72 &6.26 &5.44 &2.11 & 34 &106\\
 66764 &16:24:17.72 &-26:29:47.10 &5971 &4.08  &  1.72 &6.25 &5.40 &2.22 & 41 &106\\
 67406 &16:24:17.26 &-26:28:36.00 &5949 &4.02  &  1.72 &6.15 &5.45 &2.13 & 35 &128\\
 67707 &16:24:20.28 &-26:28:03.30 &5972 &4.07  &  1.72 &6.14 &5.37 &1.94 & 23 & 83\\
 67796 &16:24:22.25 &-26:27:51.60 &5929 &3.99  &  1.72 &6.19 &5.09 &2.10 & 34 & 94\\
 68022 &16:24:19.37 &-26:27:24.40 &5962 &4.04  &  1.72 &6.17 &4.99 &2.21 & 40 & 97\\
\hline		
\end{tabular}	
\end{center}	
{\bf Notes.} For each star, we report ID, J2000 equatorial coordinates, atmospheric
parameters, the Fe, Na and Li abundances, the Li line EWs and the spectral S/N
at the lithium line.
\end{table*}